# Using kinetic theory to examine a self-gravitating system composed of baryons and cold dark matter

**Gilberto M. Kremer**[1,a], **Martín G. Richarte**[2,3,b], **Elberth M. Schiefer**[1]

[1] Departamento de Física, Universidade Federal do Paraná, Caixa Postal 19044, Curitiba 81531-990, Brazil
[2] PPGCosmo, CCE - Universidade Federal do Espírito Santo, Vitória, ES 29075-910, Brazil
[3] Departamento de Física, Facultad de Ciencias Exactas y Naturales, Universidad de Buenos Aires, Ciudad Universitaria 1428, Pabellón I, Buenos Aires, Argentina

**Abstract** We examine the evolution of non-relativistic cold dark matter gravitationally coupled to baryons with modes deep inside the Hubble radius (sub-horizon regime) using a kinetic theory approach within the realm of Newtonian theory. We obtain the general solution for the total density perturbation and we also show that a baryon perturbation catches up with the dark matter perturbation at late times, which in turn makes possible the formation of bound structures. We extend the linear perturbation analysis by considering the turn-around event, collapse of matter, and its virialization process.

## 1 Introduction

Current lore associated with the standard cosmological model would indicate that on scales larger than 100 Mpc the universe is essentially homogeneous and isotropic, but on smaller scales there are some deviations from the mean density in the form of galaxies, galaxy clusters, amongst other configurations [1,2]. A natural question is to ask: How do such structures grow in the universe? What is the basic mechanism to aggregate matter and make it collapse in the form of a bump? The first attempt to give an answer to that question was made by Jeans long time ago [3,4]. He focused on the necessary condition under which small perturbations of a gas cloud could grow exponentially, leading to the collapse of the cloud and therefore ending in the formation of stars. In other words, the Jeans mechanism describes the gravitational instability of a self-gravitating gas cloud. Interestingly enough, there exist plenty of ways to understand the Jeans mechanism and to derive such a criterion. To be more precise, consider an initially stable and static cloud that can be initially perturbed by the environment such as a shock wave, passing spiral arms of the galaxy, etc. This configuration can collapse if the inwards directed gravitational force is bigger than the outwards directed pressure force. The critical maximal radius that allows stability depends essentially on the Newton constant, the speed of sound and and the matter density; in fact, it leads to the idea that the denser the clouds, the more unstable they become. During the collapse only a part of the gas ends up in stars, so many stars form out of one collapsing cloud, which means that young stars are born in clusters [2].

At the cosmological level, one way to get some insight in the structure formations is to look at a simplified treatment, which in this case corresponds to the Newtonian approach [2]. Indeed, the theory of Newtonian structure formation is sufficient to understand most of the processes which are well within the horizon. To do so, one must derive the full Newtonian hydrodynamics in an expanding universe and it turns out that the Boltzmann kinetic theory is the natural way to achieve such a goal [5–7].

The first step to understanding how the cooperative effects of baryons and dark matter may work in the process of structure formation is by inspecting a collisionless self-gravitating system composed of two components and then solving the coupled collisionless Boltzmann and Poisson equation together [8,9]. A system composed of baryons and dark matter leads to a total Jeans mass which is smaller than the one associated with a single component, indicating that a smaller amount of mass is needed to ignite the collapsing process. One could expect that the bumps with masses greater than the Jeans mass initiate the collapsing process, but an over-dense region in an expanding universe eventually recollapses and virialises. In the case of a single component with an expanding background it turned out that the

[a] e-mail: kremer@fisica.ufpr.br
[b] e-mail: martin@df.uba.ar

"swindle" proposal may be avoided, while the Jeans instability is expected to arise in the limit of large wavelengths [8]. Nevertheless, one must stress that the Jeans instability is not only restricted to the Newtonian (or General Relativity) realm and it can emerge within the context of alternative gravity theories as well [10–14].

This paper is organized as follows. In Sect. 2, one presents the kinetic theory formalism for dealing with a self-gravitating system of two components within the framework of Newtonian cosmology and by doing so one also examines the general conditions to achieve the Jeans instability. Besides, the perturbation of baryons and dark matter are studied along with the conditions under which the total matter starts to virialize. In Sect. 3, the conclusions are stated. We will use the metric convention $(+,-,-,-)$ and non-geometric units in which $8\pi G \neq 1$ and $c \neq 1$ unless stated otherwise.

## 2 Kinetic theory and self-gravitating components

The spacetime evolution of the one-particle distribution function $f(\mathbf{r},\mathbf{v},t)$ in the phase space spanned by the space and velocity coordinates $(\mathbf{r},\mathbf{v})$ is ruled by the Boltzmann equation. The one-particle distribution function $f(\mathbf{r},\mathbf{v},t)$ is defined by assuming that $f(\mathbf{r},\mathbf{v},t)d^3r d^3v$ gives the number of particles in the volume element $d^3r$ about the position $\mathbf{r}$ and with velocity in the range $d^3v$ about $\mathbf{v}$ at time $t$. The Boltzmann equation in the absence of collisions between the particles but in the presence of a gravitational potential $\Phi$ reads (see e.g. [8–10])

$$\partial_t f + \mathbf{v}\cdot\nabla f - \nabla\Phi\cdot\partial_{\mathbf{v}} f = 0. \tag{1}$$

On large scales the universe is spatially homogeneous and isotropic and then one can portray such a geometry using the Friedmann–Lemaître–Robertson–Walker metric [1,2]. In particular, for a spatially flat universe the line element is

$$\mathrm{d}s^2 = (c\mathrm{d}t)^2 - a(t)^2\left(\mathrm{d}x^2+\mathrm{d}y^2+\mathrm{d}z^2\right), \tag{2}$$

where $a(t)$ is the so-called cosmic scale factor. If the universe is dominated by a perfect fluid then Einstein's field equations are reduced to some coupled differential equations known as the Friedmann and acceleration equations, respectively:

$$\left(\frac{\dot a}{a}\right)^2 = \frac{8\pi G}{3}\rho, \qquad \frac{\ddot a}{a} = -\frac{4\pi G}{3}\left(\rho + 3\frac{p}{c^2}\right). \tag{3}$$

Here the dot denotes the derivative with respect to time, $G$ is the gravitational constant, while $\rho$ and $p$ are the mass density and the pressure of the source that generates the gravitational field.

In the case of a universe filled with dark matter and baryons one can treat such components as pressureless ones ($p \simeq 0$); then the mass density as a function of the cosmic scale factor is $\rho = \rho_0(a_0/a)^3$. Replacing this density in (3) leads to a power-law scale factor in terms of the cosmic time,

$$a = a_0\left(6\pi G\rho_0 t^2\right)^{\frac{1}{3}}, \tag{4}$$

where $\rho_0$ and $a_0$ describe the values of the mass density and cosmic scale factor at $t=0$, respectively.

In order to describe a system composed of dark and baryon matter subject to a gravitational field one can make use of a Boltzmann equation for each constituent plus a Poisson equation where both components are coupled gravitationally. From now on, one will consider two distributions functions, one for baryons, $f_b \equiv f(\mathbf{r},\mathbf{v}_b,t)$, and the other one for dark matter, $f_d \equiv f(\mathbf{r},\mathbf{v}_d,t)$. Both distributions satisfy the Boltzmann equation with a vanishing collisional operator (1):

$$\partial_t f_j + \mathbf{v}_j\cdot\nabla f_j - \nabla\Phi\cdot\partial_{\mathbf{v}_j} f_j = 0. \tag{5}$$

Here the index $j=\{b,d\}$ denoted baryons and dark matter, respectively. The gravitational field must fulfill the Poisson equation:

$$\nabla^2\Phi = 4\pi G\left(\int m_b f_b d\mathbf{v}_b + \int m_d f_d d\mathbf{v}_d\right) \tag{6}$$

$$= 4\pi G(\rho_b + \rho_d). \tag{7}$$

Provided these components are in equilibrium their corresponding distribution functions must be the well-known Maxwellians written in a comoving frame:

$$f_j^0(\mathbf{r},\mathbf{v}_j,t) = \frac{\rho_j}{m_j(2\pi\sigma_j^2)^{3/2}}\exp\left(-\frac{\left(\mathbf{v}_j - \frac{\dot a}{a}\mathbf{r}\right)^2}{2\sigma_j^2}\right), \tag{8}$$

where $\sigma_j$ is the dispersion velocities associated with the baryons and dark matter, respectively. Indeed, the dispersion velocities $\sigma_j$ do not depend on the space coordinates and are proportional to $1/a(t)$ for an epoch of post-recombination, namely

$$\sigma_j = \sigma_j^0\frac{a_0}{a(t)}. \tag{9}$$

One uses the possibility to relate the observationally measured coordinates $\mathbf{r}$, also known as the physical or proper coordinates, to the comoving coordinate $\mathbf{x}$ by $\mathbf{r}(t) = a(t)\mathbf{x}$. One then has the proper velocity of a given observer $\dot{\mathbf{r}}(t) = \dot a(t)\mathbf{x} + a\dot{\mathbf{x}}$; thus the second term is the so-called peculiar velocity of objects. In the absence of peculiar velocities one has the Hubble–Lamaître's law $\mathbf{v}_H = H(t)\mathbf{r}$, $H(t) = \dot a/a$ being Hubble's parameter. Thus the expansion of the universe leads to objects moving away from us at a speed proportional to the distance. Further, the Hubble parameter defines a time scale and can be used to define a length scale over which physical processes act coherently. The comoving Hubble radius is defined as $d_H = c/H$. This is also the scale at which general relativistic effects become important. For $l \ll d_H$ Newtonian gravity is often adequate [2].

Let us justify properly why the assumption of Newtonian gravity is good enough for our purpose of studying the evolution of dark matter and baryons with the help of kinetic theory. In general the process of structure formation should be studied in a fully covariant manner within the theory of General Relativity [1]. However, Newtonian gravity is generally valid at most scales of cosmological interest. It is valid in regions which are small compared to the Hubble radius $l \ll d_H$ and large compared to the Schwarzschild radius of most massive black holes. Newton equations of motion are recovered in the weak field limit of General Relativity, i.e., when $\Phi \ll c^2$. Conversely if one assumes that particle motions are non-relativistic, $v/c \ll 1$, one can reduce an arbitrary metric to the following form in the Newtonian limit [2]: $\mathrm{d}s^2 = (c^2 + 2\Phi)\mathrm{d}t^2 - \mathrm{d}\mathbf{r}^2$, with $\Phi \ll c^2$. Here $\Phi$ is the effective Newtonian potential that should be used in the equations of motion. Consider the following coordinate transformation law for a spatially flat FLRW metric:

$$\mathbf{r}(t) = a(t)\mathbf{x}, \tag{10}$$

$$\bar{t} = t - t_0 + \frac{1}{2}\frac{Ha^2x^2}{c^2} + \mathcal{O}(x^4). \tag{11}$$

In the above transformation the spatial coordinates have been changed from comoving to the proper coordinates and the time has been corrected for the gravitational redshift. The transformation changes the metric to

$$\mathrm{d}s^2 = \left(1 - \frac{\ddot{a}}{a}\frac{r^2}{c^2}\right)c^2\mathrm{d}\bar{t}^2 - \left(1 + \frac{\dot{a}^2}{a^2}\frac{r^2}{c^2}\right)\mathrm{d}r^2 - r^2 d\Omega^2. \tag{12}$$

Here one has ignored cubic and higher order terms in $r/d_H \ll 1$ because one is only interested in regions that are small compared to the Hubble radius. The coefficient in front of $\mathrm{d}r^2$ can be set to unity if $v/c \ll 1$ and $r/d_H \ll 1$. The line element (12) then reduces to $ds^2 = \left(1 - \frac{\ddot{a}}{a}\frac{r^2}{c^2}\right)c^2\mathrm{d}\bar{t}^2 - \mathrm{d}r^2 - r^2\mathrm{d}\Omega^2$. After comparing the latter expression with the metric in the Newtonian limit one can identify the effective Newtonian potential due to the homogeneous and isotropic background universe:

$$\Phi_0 = -\frac{\ddot{a}}{a}\frac{r^2}{2c^2} = \frac{2\pi}{3}G\,(\rho_b + \rho_d)\,r^2. \tag{13}$$

Equation (13) must coincide with the gravitational potential obtained from the Poisson equation, Eq. (7). The previous finding shows that the potential and the Newtonian approximation are valid only when the particle motions are non-relativistic, and the scales of interest are much smaller than the Hubble radius. Gravitational clustering occurs at scales which are much smaller than the scale of homogeneity, which current studies estimate to be 60 $\mathrm{h}^{-1}$Mpc. Such a scale is indeed much smaller than the Hubble scale, 3000 $\mathrm{h}^{-1}$Mpc. One can therefore study gravitational clustering in the Newtonian framework [2]. Summarizing the above argument, one corroborated that, deep inside the Hubble radius, Newtonian perturbation theory is in agreement with relativistic perturbation theory because the typical velocities are small compared to the speed of light. Then Newtonian and relativistic perturbation theory have to agree on the relation between single-and multi-species evolution on sub-Hubble scales [15,16]. Such a statement would confirm the results obtained from the N-body numerical simulations within the Newtonian gravity approach being almost the same (on a sub-horizon scale) as those which emerge from solving the full non-linear Einstein equation numerically [17].

One would like to examine the basic theory for structure growth in the expanding universe within the framework of kinetic theory. To do so, one introduces small perturbations in the distributions functions along with some perturbation in the gravitational potential,

$$f(\mathbf{r}, \mathbf{v}_j, t) = f_j^0\left[1 + h_j(\mathbf{r}, \mathbf{v}_j, t)\right], \tag{14}$$

$$\Phi(\mathbf{r}, t) = \Phi_0(\mathbf{r}, t) + \Phi_1(\mathbf{r}, t), \tag{15}$$

where the perturbations are assumed to be small compared with the zeroth order quantities. Replacing representations (14) and (15) into the Boltzmann and Poisson equations (5) and (7) leads to the following system of equations for $h_b$, $h_d$ and $\Phi_1$:

$$f_j^0\left[\partial_t h_j + \mathbf{v}_b \cdot \nabla h_j - \nabla\Phi_0 \cdot \partial_{\mathbf{v}_j} h_j\right] - \nabla\Phi_1 \cdot \partial_{\mathbf{v}_j} f_j^0 = 0, \tag{16}$$

$$\nabla^2\Phi_1 = 4\pi G\left(\int m_b f_b^0 h_b + \int m_d f_d^0 h_d\right) d^3v. \tag{17}$$

As is customary one neglected the products of $\nabla\Phi_1$ with $h_b$, $h_d$, $\partial_{v_b} h_b$, and $\partial_{v_d} h_d$ provided these terms are of second order in the perturbed variables. Furthermore, the perturbed variables will expand in a plane wave base where the physical wavenumber vector is $\mathbf{q}/a(t)$, while the comoving one is named $\mathbf{q}$. Thus, the factor $1/a(t)$ in the wavenumber takes into account that the wavelength is stretched out in an expanding universe:

$$h_j(\mathbf{r}, \mathbf{v}_j, t) = h_\alpha^1(\mathbf{r}, \mathbf{v}_j, t)\exp\left(i\frac{\mathbf{q}\cdot\mathbf{r}}{a(t)}\right), \tag{18}$$

$$\Phi_1(\mathbf{r}, t) = \phi(t)\exp\left(i\frac{\mathbf{q}\cdot\mathbf{r}}{a(t)}\right), \tag{19}$$

where the amplitudes $\phi$ and $h_j^1$ depend on time. In fact, the aforesaid amplitudes $h_j^1$ can be written as a linear combination of the collision invariants $\left(1, \mathbf{v}_j - H\mathbf{r}, (\mathbf{v}_j - H\mathbf{r})^2\right)$ of the Boltzmann equations:

$$h_j^1 = A_j(t) + \mathbf{B}_j(t) \cdot \left(\mathbf{v}_j - H\mathbf{r}\right) + D_\alpha(t) \left(\mathbf{v}_j - H\mathbf{r}\right)^2. \tag{20}$$

Here $A_j, \mathbf{B}_j, D_j$ are time dependent amplitudes. Now replacing (18)–(20) in (16) and (17) yields

$$\frac{\mathrm{d}A_j(t)}{\mathrm{d}t} + \left(\mathbf{v}_j - H\mathbf{r}\right) \cdot \frac{\mathrm{d}\mathbf{B}_j(t)}{\mathrm{d}t} + \left(\mathbf{v}_j - H\mathbf{r}\right)^2 \frac{\mathrm{d}D_j(t)}{\mathrm{d}t} - H\left[\mathbf{B}_j(t) + 2\left(\mathbf{v}_j - H\mathbf{r}\right) D(t)\right] \cdot \left(\mathbf{v}_j - H\mathbf{r}\right) + \left[A_j(t) + \left(\mathbf{v}_j - H\mathbf{r}\right) \cdot \mathbf{B}_j(t) + \left(\mathbf{v}_j - H\mathbf{r}\right)^2 D_j(t) + \frac{\phi(t)}{\sigma_j^2}\right] \frac{i\mathbf{q}}{a(t)} \cdot \left(\mathbf{v}_j - H\mathbf{r}\right) = 0, \tag{21}$$

$$\frac{q^2}{a^2}\phi + 4\pi G\left[\left(A_b + 3\sigma_b^2 D_b\right)\rho_b + \left(A_d + 3\sigma_d^2 D_d\right)\rho_d\right] = 0. \tag{22}$$

At this point one comment is in order. So far one has dealt with the Boltzmann equation for baryons and dark matter plus the Poisson equation for a fixed background cosmology. Nevertheless, one did not use any balance equations. Then a rather natural question to ask is: How will one recover a second order master equation for the total contrast density and for the partial contrast density of baryons and dark matter? The answer to that question is intrinsically connected to the kinetic theory in itself and how to obtain some derived physical quantities from the Boltzmann equation. In other words, one needs to calculate the perturbed balance equations and the way to obtain such first order balance equations is by multiplying Eq. (21) with the collision invariants $\left(1, \mathbf{v}_j - H\mathbf{r}, \left(\mathbf{v}_j - H\mathbf{r}\right)^2\right)$, and then integrate using some well-known results of Gaussian integrals. It turns out that one arrives at the following system of equations:

$$\frac{\mathrm{d}A_j}{\mathrm{d}t} + 3\sigma_j^2 \frac{\mathrm{d}D_j}{\mathrm{d}t} + i\frac{\sigma_j^2}{a} B_j - 6H\sigma_j^2 D_j = 0, \tag{23}$$

$$\frac{\mathrm{d}B_j}{\mathrm{d}t} + i\frac{q^2}{a}\left[A_j + 5\sigma_j^2 D_j + \frac{\phi}{\sigma_j^2}\right] - HB_j = 0, \tag{24}$$

$$3\frac{\mathrm{d}A_j}{\mathrm{d}t} + 15\sigma_j^2 \frac{\mathrm{d}D_j}{\mathrm{d}t} + i5\frac{\sigma_j^2}{a} B_j - 30H\sigma_j^2 D_j = 0. \tag{25}$$

From (23) and (25) it follows that $dA_j/dt = 0$, so one can choose $A_j = 1$ for simplicity. If one introduces the density contrasts for each component as $\bar{\delta}_{\rho_j} = \int m h_j d^3v/\rho_j = A_j + 3\sigma_j^2 D_j$, and then combines (23) and (24) along with the Poisson equation, Eq. (22), one gets a coupled system of differential equations for the density contrasts which reads

$$\ddot{\delta}_{\rho_j} + 2H\dot{\delta}_{\rho_j} + \frac{5\sigma_j^2 q^2}{3a^2}\delta_{\rho_j} - 4\pi G\left(\sum_j \rho_j \delta_{\rho_j} + \frac{2}{5}\sum_j \rho_j\right) = 0, \tag{26}$$

where it was useful to recast Eq. (26) in a more canonical way by introducing a shift in the contrast density as follows: $\delta_{\rho_j} = \bar{\delta}_{\rho_j} - 2/5$.

To identify the Jeans wavenumber for the total system (baryon plus dark matter) it requires one to introduce the total mass density $\rho_t = \sum_j \rho_j$ along with the total density contrast $\rho_t\delta_t = \sum_j \rho_j\delta_{\rho_j}$. As a result of multiplying (26) by $\rho_j/\rho_t$ and then summing over the two species we find that the time evolution of the total density contrast can be written as

$$\ddot{\delta}_t + 2H\dot{\delta}_t + \sum_j\left(\frac{5\rho_j\sigma_j^2 q^2}{3a^2\rho_t}\delta_{\rho_j}\right) = 4\pi G\rho_t\left(\delta_t + \frac{2}{5}\right). \tag{27}$$

As is well known the total adiabatic sound speed can be recast as $v_s = \sqrt{5\sigma^2/3}$, while for a mixture of two components it reads $v_s = \sqrt{\sum_j \rho_j v_{sj}/\rho_t} = \sqrt{\sum_j 5\rho_j\sigma_j^2/(3\rho_t)}$. Hence for the third term on the left-hand side of (27) we could introduce a mean speed of sound $\bar{v}_s$ through $\sum_j 5\rho_j\sigma_j^2\delta_{\rho_j}/(3\rho_t) = \bar{v}_s^2\delta_t$, so that

$$\ddot{\delta}_t + 2H\dot{\delta}_t + \left(\frac{\bar{v}_s^2 q^2}{a^2} - 4\pi G\rho_t\right)\delta_t = \frac{8}{5}\pi G\rho_t. \tag{28}$$

In order to see that the relations $\rho_t\delta_t = \sum_j \rho_j\delta_{\rho_j}$ and $\sum_j 5\rho_j\sigma_j^2\delta_{\rho_j}/(3\rho_t) = \bar{v}_s^2\delta_t$ are consistent between themselves, one should have $5\rho_j\sigma_j^2/(3\rho_t\bar{v}_s^2) = \rho_j/\rho_t$, which leads to the restriction $v_{sj}^2 = 5\sigma_j^2/3 = \bar{v}_s^2, \forall j$ and the identification $\bar{v}_s \equiv v_s$.

Here the source term corresponds to the total density and does not involve the perturbed density contrast. Equation (28) tells us that a dissipative effect enters through the usual friction term proportional to $2H$. One must emphasize that the physical Jeans scale (length or wavelength) is obtained by demanding that the term $\delta_t$ vanishes, namely, the Jeans wavenumber is $q_J = (\sqrt{3/2})Ha/v_s$. The Jeans length can heuristically be derived by balancing the sound crossing time, $t_s \propto a/v_s q_J$, with the gravitational free-fall time, $t_{\rm ff} \propto 1/\sqrt{G\rho_t}$, which yields the same result as mentioned above in an intuitive manner. Moreover, this scale seems to be sensitive to the thermal dispersion velocity of the particles through $v_s$ and the total material content $\rho_t$; notice that one considers the universe after decoupling, then one can focus on dark matter and baryons only. One should mention also that the Jeans wavenumber for each component in principle is different provided the propagation speeds are not the same. Here one is looking at the Jeans length of the composed system. In this context, the comoving Jeans wavenumber can be written in terms of the scale factor as $q_J = (4\pi G\rho_t^0 a_0)^{1/2} a^{1/2}/v_s^0$—by considering that $v_s = v_s^0(a_0/a)$ and $\rho_t = \rho_t^0(a_0/a)^3$—while the comoving Jeans length is simply $\lambda_J = 2\pi/q_J$. One can prove that perturbations can grow for $q \le q_J$; otherwise they just oscillate ( $q \ge q_J$). The latter result is consistent with the behavior of the Jeans scale in a universe dominated by matter after decoupling time, $q_J \propto a^{1/2}$ [1]. One final comment: the fact that $\sigma_j = \sigma_j^0 a^{-1}$ (which is equivalent to having a pressure $p_j = \sigma_j^2 \rho_j \propto a^{-5}$) implies that the collisionless fluid may be treated as pressureless as long as $q < q_J$. As one is trying to arrive at the general solution without demanding the latter condition the system of equations will be much harder to solve than the usual case. One will examine the situation where the condition $q > q_J$ holds as well.

In order to carry on it is mandatory to find a way to solve the above equation. One route is to make explicit the Hubble factor:

$$\ddot{\delta}_t + 2H\dot{\delta}_t + \frac{3}{2}H^2\left(\frac{q^2}{q_J^2} - 1\right)\delta_t = \frac{3}{5}H^2. \tag{29}$$

A more useful expression than the one given by (29) can be obtained by performing a change of variables and introducing a new time, $\eta = (2/3)\ln(H_0 t)$, and a constant Jeans wavenumber, $q_J^0 = (4\pi G\rho_t^0)^{1/2} a_0/v_s^0$; in this way one is able to get rid of the $H^2$ factor, yielding

$$\frac{\mathrm{d}^2\delta_t}{\mathrm{d}\eta^2} + \frac{1}{2}\frac{\mathrm{d}\delta_t}{\mathrm{d}\eta} + \left[\left(\frac{3}{2}\right)^{\frac{1}{3}}\left(\frac{q}{q_J^0}\right)^2 \frac{1}{e^\eta} - \frac{3}{2}\right]\delta_t = \frac{3}{5}. \tag{30}$$

Let us analyze what kinds of limiting regimes follow from (30). For large values of the wavelength with respect to the Jeans wavelength, $q/q_J^0 = \lambda_J^0/\lambda \ll 1$, Eq. (30) can be approximated by

$$\frac{\mathrm{d}^2\delta_t}{\mathrm{d}\eta^2} + \frac{1}{2}\frac{\mathrm{d}\delta_t}{\mathrm{d}\eta} - \frac{3}{2}\delta_t = \frac{3}{5}, \tag{31}$$

whose solution as a function of the usual cosmic time is given by

$$\delta_t(t) = C_1\,(H_0 t)^{\frac{2}{3}} + \frac{C_2}{H_0 t} - \frac{2}{5}, \tag{32}$$

where $C_1$ and $C_2$ are integration constants. This solution represents the well-known growth of the total density contrast for a matter-dominated universe proportional to $t^{\frac{2}{3}}$ and it corresponds to Jeans instability.

Equation (30) can be approximated in the limiting case of small values of the wavelength with respect to the Jeans wavelength $q/q_J^0 = \lambda_J^0/\lambda \gg 1$ by the expression

$$\frac{\mathrm{d}^2\delta_t}{\mathrm{d}\eta^2} + \frac{1}{2}\frac{\mathrm{d}\delta_t}{\mathrm{d}\eta} + \left(\frac{3}{2}\right)^{\frac{1}{3}}\left(\frac{q}{q_J^0}\right)^2 \frac{1}{e^\eta}\delta_t = \frac{3}{5}, \tag{33}$$

and the solution of Eq. (33) in terms of the usual cosmic time admits a closed form,

$$\begin{aligned}\delta_t(t) = {} & \cos\left[\left(\frac{4\sqrt{6}}{H_0 t}\right)^{\frac{1}{3}}\frac{q}{q_J^0}\right] \\ & \times\left\{A_1 - \frac{12}{5}\mathrm{Ci}\left[\left(\frac{4\sqrt{6}}{H_0 t}\right)^{\frac{1}{3}}\frac{q}{q_J^0}\right]\right\} \\ & - \sin\left[\left(\frac{4\sqrt{6}}{H_0 t}\right)^{\frac{1}{3}}\frac{q}{q_J^0}\right] \\ & \times\left\{A_2 + \frac{12}{5}\mathrm{Si}\left[\left(\frac{4\sqrt{6}}{H_0 t}\right)^{\frac{1}{3}}\frac{q}{q_J^0}\right]\right\}.\end{aligned} \tag{34}$$

Here $A_1$ and $A_2$ are integration constants, while $\mathrm{Ci}(x) = -\int_x^\infty \mathrm{d}t\,(\cos t/t)$ and $\mathrm{Si}(x) = \int_0^x \mathrm{d}t\,(\sin t/t)$ represent the cosine and sine integrals, respectively. We infer from Eq. (34) that it represents oscillations of the total density contrast with respect to time $t$ for values of $\left(\frac{4\sqrt{6}}{H_0 t}\right)^{\frac{1}{3}}\frac{q}{q_J^0}$ not too small. When the time increases, the factor $\left(\frac{4\sqrt{6}}{H_0 t}\right)^{\frac{1}{3}}\frac{q}{q_J^0}$ becomes much smaller than unity and the total density contrast will grow provided that the cosine integral term will increase for large values of the cosmic time.

In order to gain more insight in the behavior of the total density contrast with respect to cosmic time we introduce another dimensionless time, called $\tau = (H_0 t)$, and write (29) as follows:

$$\tau^2\delta_t'' + \frac{4}{3}\tau\delta_t' + \frac{2}{3}\left[\left(\frac{2}{3\tau}\right)^{\frac{2}{3}}\left(\frac{q}{q_J^0}\right)^2 - 1\right]\delta_t = \frac{4}{15}, \tag{35}$$

where the prime refers to differentiation with respect to $\tau$. We solved (35) for the initial conditions $\delta_t(10^{-3}) = 0.1$ and $\delta_t'(10^{-3}) = 0$ (say) in the limit cases of small and large wavelengths. The aforesaid numerical simulations are displayed in Figs. 1 and 2. Figure 1 tells us that for small wavelengths with respect to the Jeans wavelength, $\lambda_J^0/\lambda = 50$, the total density contrast exhibits some small oscillations for small values of time $\tau$ but eventually as $\tau$ increases the total density contrast begins to grow. In the opposite limit, with $\lambda_J^0/\lambda = 0.5$ we find that there is only exponential growth of the total density contrast as can be seen from Fig. 2.

The next task to tackle is to solve the master equation, Eq. (26), for each species, at least under some reasonable conditions. Equation (26) can be written as

$$\tau^2 \delta_j'' + \frac{4}{3}\tau \delta_j' + \frac{2}{3}\left[\left(\frac{2}{3\tau}\right)^{\frac{2}{3}}\left(\frac{q}{q_J^0}\right)^2\left(\frac{v_{sj}^0}{v_s^0}\right)^2 - \sum_j \delta_j \frac{\rho_j^0}{\rho_t^0}\right] = \frac{4}{15}, \tag{36}$$

where $\tau = H_0 t$, $\delta_{\rho_j} \equiv \delta_j$ and we used the relationships $v_s = v_s^0(a_0/a)$, $v_j = v_j^0(a_0/a)$, $\rho_t = \rho_t^0(a_0/a)^3$ and $\rho_j = \rho_j^0(a_0/a)^3$. For some given ratios $\rho_j^0/\rho_t^0$, $v_{sj}^0/v_s^0$, and $q/q_{J0}$ one can solve the coupled system of differential equations (37) numerically. As we have done in our previous analysis we will select the same values for the ratio $q/q_J^0 = \lambda_J^0/\lambda$ in the case of small wavelengths in relation to the Jeans scale ($\lambda_J^0/\lambda = 50$) and in the case of large wavelength $\lambda_J^0/\lambda = 0.5$. Accordingly, we begin by recalling that the mass density ratio $\rho_d^0/\rho_b^0$ could be associated with the density parameter ratio $\Omega_d^0/\Omega_b^0$, which is about 5.5 [18], i.e., $\rho_j^0/\rho_t^0 = \Omega_d^0/\Omega_b^0 \approx 5.5$. Perhaps not surprisingly, the ratio of the sound speeds $v_{sj}^0/v_s^0$ is not usually fixed in the literature. However, we will follow [8,9] and use the value $v_{sd}^0/v_{sb}^0 = 1.83$, which was inferred from the simulations with Maxwellian distributions for Milky Way-like galaxies with baryonic and dark matter [19]. Carrying on, once the parameters are fixed the coupled system of the differential equation (37) can be solved numerically. We set the initial conditions $\delta_d(10^{-3}) = 0.1$, $\delta_b(10^{-3}) = 0$ and $\delta_d'(10^{-3}) = \delta_b'(10^{-3}) = 0$; that is, at the very beginning only dark matter has a non-vanishing contribution. Figures 3 and 4 display the baryon density contrast $\delta_b$ and the dark matter density contrast $\delta_d$ in terms of $\tau = H_0 t$, the cosmic time. Figure 3 shows that, for small wavelengths with respect to the Jeans wavelength, $\lambda_J^0/\lambda = 50$, both contrast densities exhibit some oscillations for small values of $\tau$, but as time proceeds they start to grow. For large wavelengths with respect to the Jeans wavelength, $\lambda_J^0/\lambda = 0.5$, we find that both density contrasts obey exponential growth, implying the development of the Jeans instability [cf. Fig. 4]. Indeed, we note that due to the imposed initial conditions the baryons density contrast starts from zero but at later times it becomes equal to the dark matter density contrast, since the slope of the baryon density contrast is more accentuated than the dark matter ones.

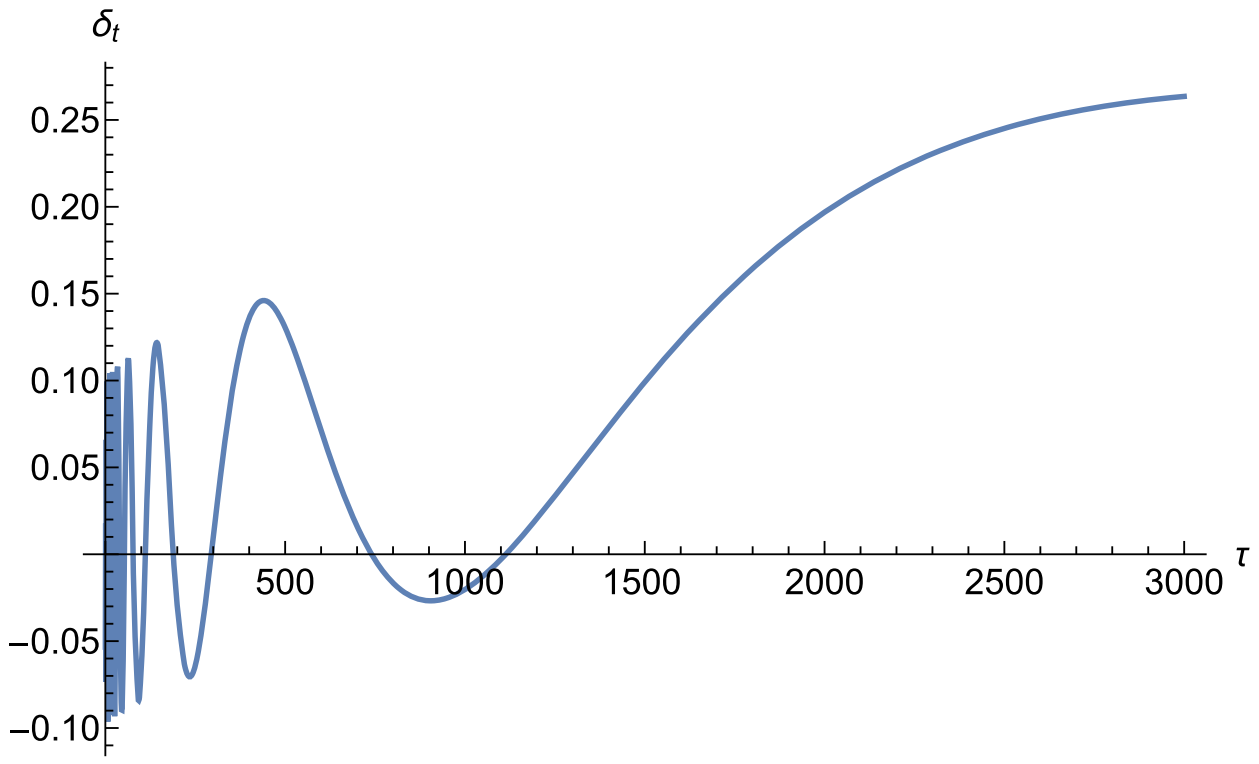

**Fig. 1** Total density contrast $\delta_t$ as a function of time $\tau = H_0 t$ for small wavelength with respect to Jeans' wavelength $\lambda_J^0/\lambda = 50$

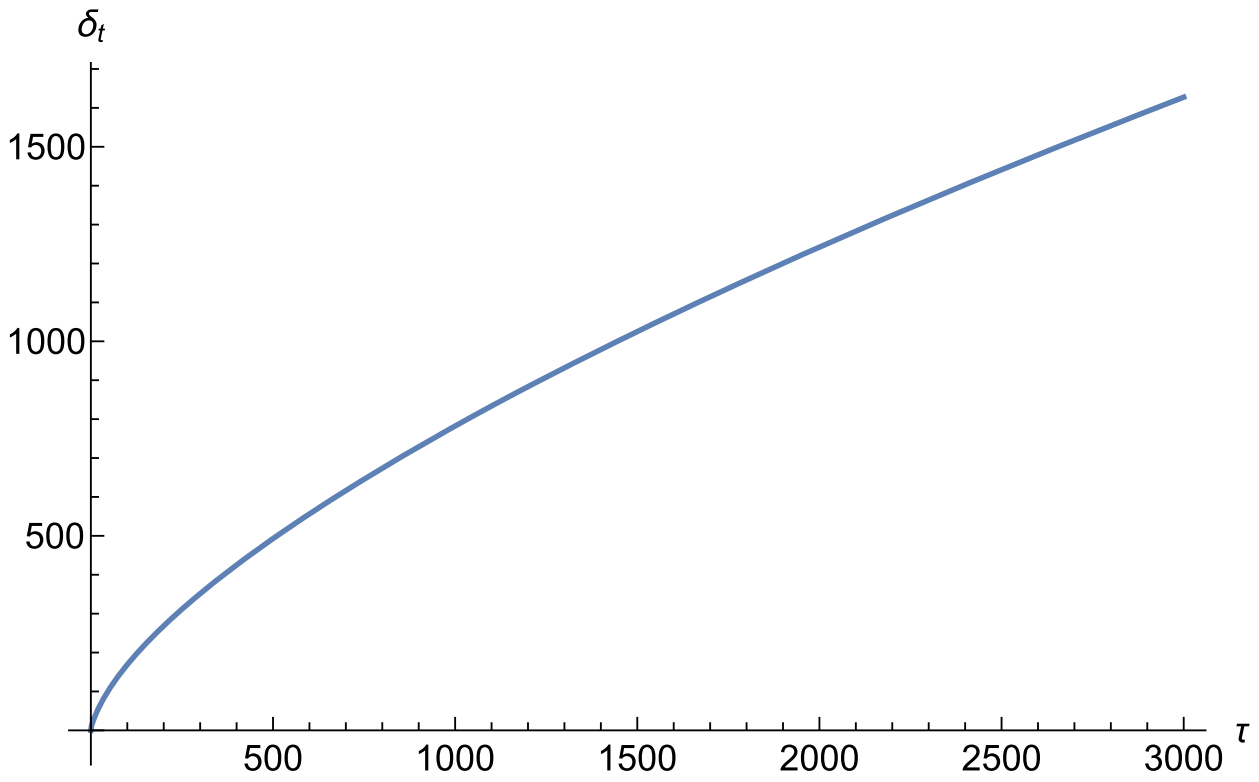

**Fig. 2** Total density contrast $\delta_t$ as a function of time $\tau = H_0 t$ for large wavelength with respect to Jeans' wavelength $\lambda_J^0/\lambda = 0.5$

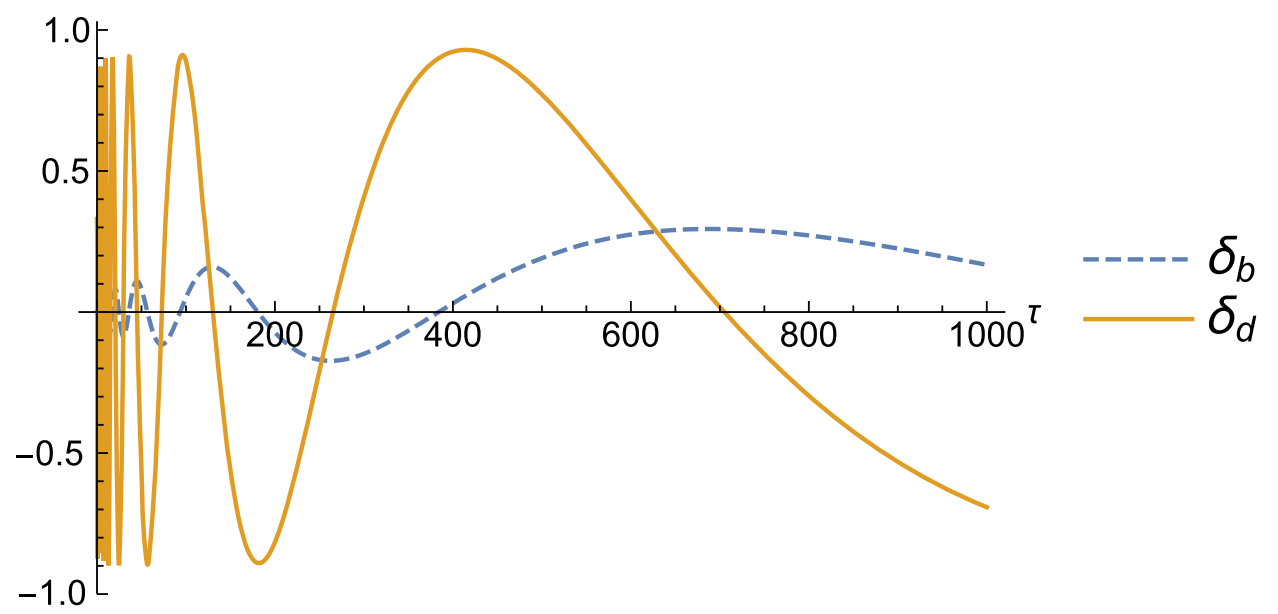

**Fig. 3** Density contrasts of baryons $\delta_b$ and dark matter $\delta_d$ as functions of time $\tau = H_0 t$ for small wavelength with respect to Jeans' wavelength $\lambda_J^0/\lambda = 50$

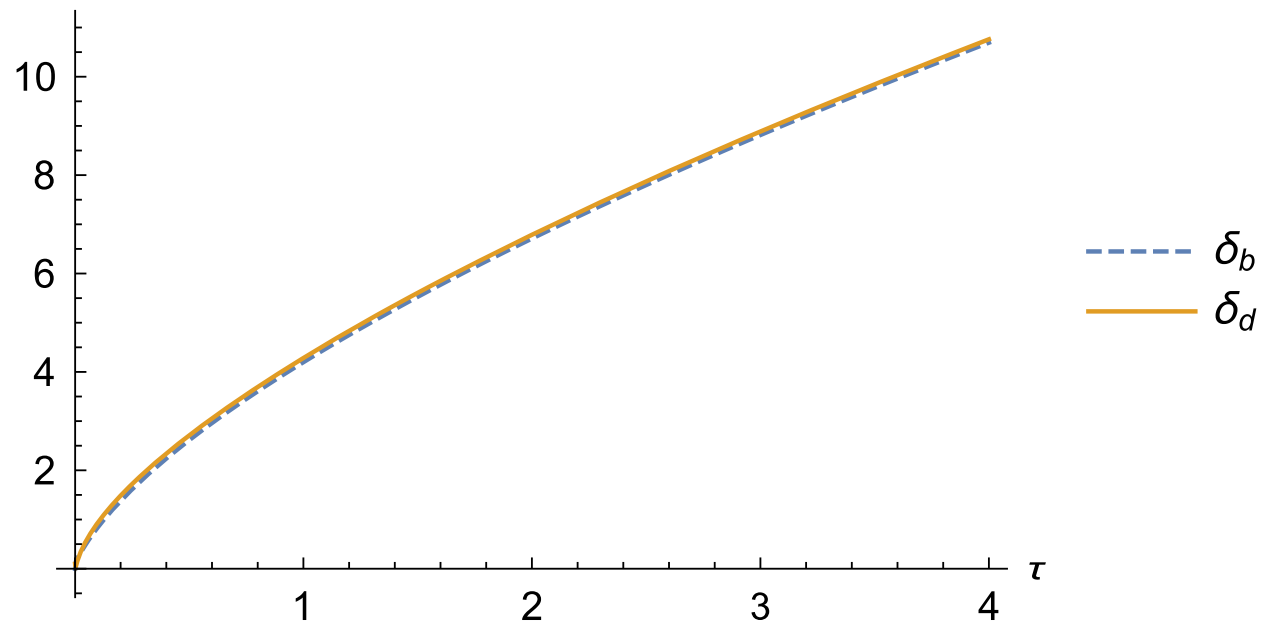

**Fig. 4** Density contrasts of baryons $\delta_b$ and dark matter $\delta_d$ as functions of time $\tau = H_0 t$ for large wavelength with respect to Jeans' wavelength $\lambda_J^0/\lambda = 0.5$

In characterizing the properties of the composed system made of dark matter and baryons and how they evolve with time, we are certainly interested in double-checking the previous numerical analysis by solving analytically the set of equations in terms of the cosmic time variable $\eta = (2/3)\ln(H_0 t)$. As such, it is important to notice that the abundance of the two components is encoded in the master equation (29), which reads

$$\frac{d^2\delta_j}{d\eta^2} + \frac{1}{2}\frac{d\delta_j}{d\eta} + \left(\frac{3}{2}\right)^{\frac{1}{3}}\left(\frac{q}{q_J^0}\right)^2 \frac{1}{e^\eta}\left(\frac{v_{sj}^0}{v_s^0}\right)^2 \delta_j = \frac{3}{2}\sum_j \delta_j \frac{\rho_j^0}{\rho_t^0} + \frac{3}{5}. \quad (37)$$

We need to corroborate that Eq. (37) leads to the correct physics; otherwise the current formalism has no merit at all. In particular, one has to ensure that baryons can catch up with dark matter perturbations after recombination [2]. Of course, baryons only contribute a small amount to the total matter density, but all the visible structures in the universe are made of them. One explores the simplest case in which the modes satisfy the condition $q \ll q_J$ for both baryons and dark matter. Although both components act as sources to the gravitational potential perturbation, one will assume that dark matter dominates the background density so the perturbed Poisson equation will have only as a source the perturbed dark matter density. In fact, one also will neglect in Eq. (37) the term proportional to $\delta_j$ provided it appears multiplied by the factor $(q/q_J)^2$, which is considerably small in this regime. The aforesaid approximations yield the following system of equations for dark matter and baryons:

$$\frac{d^2\delta_d}{d\eta^2} + \frac{1}{2}\frac{d\delta_d}{d\eta} \simeq \left(\frac{3}{5} + \frac{3}{2}\delta_d f_{d0}\right), \quad (38)$$

$$\frac{d^2\delta_b}{d\eta^2} + \frac{1}{2}\frac{d\delta_b}{d\eta} \simeq \left(\frac{3}{5} + \frac{3}{2}\delta_d f_{d0}\right), \quad (39)$$

where $f_{d0} = \rho_d^0/\rho_t^0$. The general solution for the evolution of the dark matter perturbations is given by the superposition of three terms, namely, a constant mode, a decay mode and a growing mode:

$$\delta_d = \delta_{d0} + \delta_- e^{\lambda_- \eta} + \delta_+ e^{\lambda_+ \eta}, \quad (40)$$

where the exponents are $\lambda_\pm = -\frac{1}{4} \pm \sqrt{\frac{1}{16} + \frac{3}{2} f_{d0}}$ while $\delta_{d0}$ and $\delta_\pm$ are some constants. Given the fact that the growing mode is the relevant one at late times it will be enough to consider that $\delta_d \simeq \delta_+ e^{\lambda_+ \eta}$. Note that the growing mode of the dark matter perturbations provides the driving term in the master equation for baryons. Armed with this new fact, one can return to the task of finding the behavior of baryons which it involves to provide some initial conditions. One will assume that at early times, nearly the recombination era, $\delta_b(\eta_i) = \delta_b'(\eta_i) \simeq 0$. Then the general solution for baryons reduces to

$$\delta_b = \delta_{d0}\left[-(2\lambda_+ + 1) + 2\lambda_+ e^{\frac{(\eta_i - \eta)}{2}} + e^{\frac{\lambda_+(\eta - \eta_i)}{2}}\right]. \quad (41)$$

Equation (41) tells that at very early times $\delta_b \simeq 2\lambda_+ \delta_{d0} e^{-\eta/2}$, while in the opposite limit, at late times, $\delta_b \simeq \overline{\delta}_{d0} e^{\lambda_+ \eta} = \delta_d$. So the baryon perturbation catches up with the dark matter perturbation at late times. This is another reason why we need dark matter. Without dark matter to set up gravitational potential wells for the baryons to fall into, it would be hard to explain how we can have bound baryon structures today, given that $\delta_b(\eta_i)$ is of the order of $10^{-5}$. Thus, baryon perturbations would still be in the linear regime today without the help of dark matter perturbations, making bound structures like galaxies and clusters of galaxies impossible. Nevertheless, for a better understanding of the clustering properties of a system composed of dark matter plus baryons we will need to extend the analysis to a scenario where the Boltzmann equation is solved in curved spacetime for the composed system and the fully relativistic corrections of General Relativity are taken into account as well [20].

The next step of consistency is to explore the non-linear regime of total density perturbation. Suppose now, as a toy model for the formation of non-linear (gravitationally bound) structures, one has a spherical over-density of radius $R$ and mass $M$ embedded into the otherwise homogeneous spatially flat and matter-dominated universe (EdS). Given the fact that it is over-dense, this configuration will reach a maximum radius and subsequently contract until collapse. Such a toy model is a reasonable approximation provided the distribution of the dark matter in the universe can be considered as composed of individual so-called halos, approximately spherical over-dense clouds of dark matter which can reach highly non-linear densities in their centers. This means that one can work with a Newtonian equation of motion for the radius,

$$\ddot{R} = -\frac{GM}{R^2}, \quad (42)$$

where the mass of the halo can be expressed in terms of the turn-around radius and the turn-around density as $M = \frac{4\pi}{3} R_{\rm ta}^3 \rho_{\rm ta}$. The density at the turn-around event involves the critical density of the background and the over-density $\zeta$ of the halo with respect to the background at turn-around, thus $\rho_{\rm ta} = \zeta\rho_c(a_{\rm ta}) = (3H_{\rm ta}^2/8\pi G)\zeta$. It is useful to introduce some more convenient variables by parameterizing all the physical quantities in terms of variables evaluated at the turn-around event, that is to say, $R = R_{\rm ta} y$, $a = x a_{\rm ta}$ and $\tau = H_{\rm ta} t$ with $H_{\rm ta} = H_0 a_{\rm ta}^{3/2}$. Now, the Newton equation for the spherical halo reads

$$y'' = -\frac{\zeta}{2y^2}. \quad (43)$$

In order to solve (48) one could consider the case where initially the halo has zero radius and then reaches its maximum radius at the turn-around event. These assumptions imply that the boundary conditions for the shell must be: $y'(x = 1) = 0$

and $y(x=0)=0$. After integrating once and imposing the boundary condition $y'(x=1)=0$ one arrives at the velocity of the shell in terms of its radius, $y'=\pm\sqrt{\zeta}(y^{-1}-1)^{1/2}$; the minus stands for the evolution of the shell before the turn-around event and the plus sign is applied after the turn-around. Using the Friedmann equation one can get a parametric solution $\tau=\tau[y]$:

$$\tau[y,\zeta]=\frac{1}{\sqrt{\zeta}}\left(\frac{1}{2}\arcsin(2y-1)-\sqrt{y-y^2}+\frac{\pi}{4}\right). \tag{44}$$

The shell must be able to collapse and by doing so it must form a structure, so the expansion of the shell must come to a halt and the radial velocity must vanish at a finite time. This stage of zero radial velocity is the so-called turn-around event where the perturbation reaches its maximum radius. For the case under study such a situation occurs at $x=1=y$ and $\tau_{\rm ta}=2/3$, which in turn implies that the over-density of the halo is $\zeta=(3\pi/4)^2$. Furthermore, the collapse time is reached at $\tau_{\rm coll}=2\tau_{\rm ta}=4/3$ for $x_{\rm coll}=\sqrt[3]{4}$.

To obtain the parameters that characterize the collapsing phase, one begins with the behavior of $\tau$ at early times. Making an expansion of $\tau[y,\zeta]$ at the lowest order possible in $y$ around $y=0$ yields $\tau\simeq(8/9\pi)y^{2/3}(1+3y/10)$. Furthermore, one defines the over-density inside the halo in relation to the background density as $\Delta\equiv\rho_{\rm halo}\zeta/\rho_{\rm bg}=\zeta(x/y)^3$. Replacing the expansion for $\tau$ into the definition of over-density inside the halo leads to $\Delta=1+3y/5$. Then the linear density contrast inside the halo is $\delta=\Delta-1=3y/5$. By extrapolating this formula linearly until the turn-around event one gets $\delta_{\rm ta}=\delta(y)/x\simeq 3y/5x$ but $x[\tau(y)]\simeq\zeta^{-1/3}y$ so the contrast linear density at the turn-around is $\delta_{\rm ta}\simeq(3/5)\zeta^{-1/3}\simeq 1.06$. In the subsequent phase after the turn-around, when the collapse is reached, the inner contrast linear density is $\delta_{\rm coll}\simeq x_{\rm coll}(3/5)\zeta^{-1/3}\simeq 1.69$. In other words, the halo of dark matter has already collapsed when its expected linear density reaches the value $\delta_{\rm coll}\simeq 1.69$. Notice that the previous result is independent of the mass M, the initial over-density, and the epoch of virialization. The next step in the evolution is to consider what happens when the halo reached the virial equilibrium: essentially the potential energy of the halo must be twice that at the turn-around event and its radius must decrease at $y_{\rm v}=1/2$, implying that $\Delta_{\rm v}=(2x_{\rm coll}/)^3\zeta\simeq 178$. Therefore, a halo in virial equilibrium is expected to have a mean density nearly 178 times higher than the background although in numerical simulations the density contrast is fixed at the value 200, furnishing in this way a natural definition of the virial radius of a virialized object [2]. The lesson from this simple analysis is that density perturbations can form bound structures generated by gravitational collapse after they become 200 times as dense as the background. Such a result seems to be consistent with the full results from N-body simulations where galaxies and clusters of galaxies separate out as distinct gravitationally bound structures when their densities are at least 100 times greater than the background density [2]. However, this must be thought of just as a heuristic rule provided one is using the fact that the linear theory is valid until $\Delta_{\rm coll}\geq 1$, where the process of virialization cannot be stopped.

Having mentioned the ideal picture of halos, one must also say that the halo is not necessarily isolated from the background and there will take place a constant inflow of material into the halo, or the halo might even merge with another halo. Thus, the evolution of a halo after its formation is quite non-trivial so it cannot be easily analyzed within this simplistic model. However, if one assumes that the continuous inflow of material and the merging with other halos produce new halos which are still characterized by the same density ratio $\Delta_{\rm v}=(2x_{\rm coll}/)^3\zeta\simeq 178$, then one would expect that the mean density of a typical halo with a given mass M scales with redshift like $\rho_{\rm v}=\Delta_{\rm v}\rho_{\rm bg}\propto(1+z)^3$, whereas its physical radius should go as $r_{\rm v}\propto\rho_{\rm v}^{-1/3}$. Notice that the over-density does not depend on the mass of the perturbation, on the initial over-density, nor on the epoch of virialization $t_{\rm v}$. Thus, whenever one observes an over-density of the order of $\delta_{\rm v}$, one positively assumes that the corresponding structure is virialized (or close to virialization) irrespective of its mass or formation history.

Last but not least, it is physically meaningful to devote some efforts to understand how the virialization process takes places in more detail. The main goal will be to estimate the redshift for which the virialization occurs within the context of an EdS expanding universe. As mentioned before, the virialised density (inside the halo) must be at least 100 bigger than the background density (criterion):

$$\rho_{\rm v}\geq 100\times 3H_0^2\frac{\Omega^0_{\rm bg}}{8\pi G}(1+z_{\rm v})^3. \tag{45}$$

If the particles inside the virialized region have a velocity dispersion $\sigma_{\rm d}$, the virial theorem states that $\langle V\rangle=-2\langle K\rangle$, $V$ being the gravitational potential, $K$ the kinetic energy, and the brackets indicate the average value. From the latter fact, one gets $M_{\rm v}\sigma_{\rm d}^2=GM_{\rm v}^2/R_{\rm v}$ and therefore the radius is $R_{\rm v}=GM_{\rm v}/\sigma_{\rm d}^2$. Inserting the latter result into Eq. (45) one obtains

$$(1+z_{\rm v})\leq 0.47\left(\frac{\sigma_{\rm d}}{100\,{\rm kms}^{-1}}\right)^2\left(\frac{M_{\rm v}}{10^{12}M_\odot}\right)^{-\frac{2}{3}}(\Omega^0_{\rm bg}h^2)^{-\frac{1}{3}}. \tag{46}$$

Taking as fiducial value $\Omega^0_{\rm bg}h^2=0.11$ one gets a more or less heuristic result,

$$(1+z_{\rm v})\leq 0.93\left(\frac{\sigma_{\rm d}}{100\,{\rm kms}^{-1}}\right)^2\left(\frac{M_{\rm v}}{10^{12}M_\odot}\right)^{-\frac{2}{3}}. \tag{47}$$

Equation (47) tells us that for a typical galaxy as the Milky Way with $\sigma_{\rm d}=300\,{\rm kms}^{-1}$ and a total virialized mass of $M_{\rm v} = 10^{12} M_{\odot}$, the virialized redshift should be around $z_{\rm v} \leq 7$, so clusters of galaxies must have formed not too long ago in terms of the total evolution of the universe.

Given the fact that our universe is currently accelerating it seems interesting to explore how our analysis is modified once a dark energy component (a cosmological constant) is taken into account. To do so, we proceed as before by assuming that we have a sphere (halo) of radius $R$ enclosing a mass $m$ and its dynamic is governed by Newton equation:

$$\ddot{R} = -\frac{Gm}{R^2} + \frac{\Lambda}{3} R, \tag{48}$$

where $\Lambda$ is a cosmological constant term. Notice that the cosmological constant appears provided the background cosmology has changed and the Friedmann equation now reads $3H^2 = 8\pi G(\rho_b + \rho_d + \rho_x)$ with $\rho_x = \Lambda$. Equation (48) has as a first integral the energy conservation equation (per unit mass),

$$\mathcal{E} = \frac{E}{m} = \frac{\dot{R}^2}{2} - \frac{Gm}{R} - \frac{\Lambda}{6} R^2. \tag{49}$$

The total energy contained in the shell of radius $R$ can be obtained by integrating Eq. (49). In doing so, we consider that the differential mass is $\mathrm{d}m = \rho 4\pi r^2 \mathrm{d}r$; then the total mass becomes $M = 4\pi R^3 \rho/3$ when the total density is constant.

We would like to determine how the cosmological constant affects the virialization radius. In fact, we could expect that the inclusion of the cosmological constant would potentially affect the turn-around radius and therefore the virialization radius as well. To prove this, it is convenient to compute the total potential energy contained in a sphere of constant density using Eq. (49). It reads

$$V = -\frac{3GM^2}{5R} - \frac{\Lambda}{10} M R^2. \tag{50}$$

The virial theorem states that $\langle K \rangle = \sum_{j=g,cc} \frac{n_j}{2} \langle V_j \rangle$ for $V_j \propto R^{n_j}$, where $n_g = -1$ for the purely gravitational potential and $n_{cc} = 2$ corresponds to the potential energy associated with the cosmological constant. The interesting point as regards the theorem is that it allows us to replaced the average kinetic energy by the average potential energy in the energy conservation law. We need to compare the turn-around event with zero kinetic energy with the state of the sphere at the virialized radius. At the turn-around event the total energy is $E_{\rm ta} = -\frac{3GM^2}{5R_{\rm ta}} - \frac{\Lambda}{10} M R_{\rm ta}^2$, while at the virialized radius we obtained $E_{\rm v} = 2\langle V_{cc} \rangle + \langle V_g \rangle/2$. We define the two dimensionless variables $x = R_{\rm v}/R_{\rm ta}$ and $\vartheta = \Lambda R_{\rm ta}^3/3GM$. The energy conservation equation ($E_{\rm ta} = E_{\rm v}$) allows us to obtain a cubic equation for $x = R_{\rm v}/R_{\rm ta}$ in order to estimate the effects introduced by $\Lambda$:

$$2\vartheta x^3 - (2 + \vartheta)x + 1 = 0. \tag{51}$$

In the case without cosmological constant we basically have $x = 1/2$; however, we expect that the ratio $R_{\rm v}/R_{\rm ta}$ should be less than $1/2$ provided the cosmological constant contributes the total energy in the form of $V_{cc} = -\frac{\Lambda}{10} M R^2$. To confirm our reasoning, it is convenient to write the ratio $R_{\rm v}/R_{\rm ta}$ as a small deviation of $1/2$, namely $x = 1/2 + \delta$ with $\delta \ll 1$. Replacing $x = 1/2 + \delta$ in (51) and expanding at first order in $\delta$ we are able to write $\delta$ in terms of $\vartheta$:

$$\delta = \frac{\frac{\vartheta}{4}}{-2 + \frac{\vartheta}{2}}. \tag{52}$$

Substituting (52) in the definition of $x$ we find that the ratio $R_{\rm v}/R_{\rm ta}$ is given by

$$x = \frac{1 - \frac{\vartheta}{2}}{2 - \frac{\vartheta}{2}}, \tag{53}$$

which gives $x < 1/2$ for $0 < \vartheta < 1$ as a general result. In fact, we can go further by considering that $\vartheta \ll 1$ and ending with $x \simeq 1/2(1 - \vartheta/4) < 1/2$. We can conclude that the virialized radius is smaller if the dark energy component of the universe is taken into account. In a way, we can say that the cosmological constant term in the total potential energy helps the system to reach an equilibrium configuration much faster. $N$-body numerical simulations show that $R_{\rm v} \simeq 0.483 R_{\rm ta}$ for a non-vanishing $\Lambda$ [21].

In order to guarantee that the system has virialized we must check that the average density inside the halo must be much bigger than the critical density of the background,

$$\rho_{\rm v} = \frac{4M_{\rm v}}{3\pi R_{\rm v}^3} \geq \zeta \frac{3H_0^2}{8\pi G} \left( \Omega_{\rm bg}^0 (1 + z_{\rm v})^3 + \Omega_\Lambda^0 \right), \tag{54}$$

with $\Omega_\Lambda^0 = 1 - \Omega_{\rm bg}^0$ and $\zeta \simeq 100$. We would like to obtain a linear relation between $\sigma_{\rm d}^2$ and $R_{\rm v}$; however, the inclusion of the cosmological constant spoils that possibility. Indeed, the virial theorem leads to a relation between $\sigma_{\rm d}^2$ and $R_{\rm v}$ that is non-linear:

$$\sigma_{\rm d}^2 = \frac{3GM_{\rm v}}{5R_{\rm v}} \left( 1 - \frac{\Lambda R_{\rm v}^3}{3GM_{\rm v}} \right). \tag{55}$$

Instead of determining $z_{\rm v}$ we can consider that $z_{\rm v}$ is fixed and derive the typical order of magnitude for the virial radius with the help of (54),

$$R_{\rm v} \leq M_{\rm v}^{\frac{1}{3}} \left( \frac{2G}{\zeta H_0^2} \right)^{\frac{1}{3}} \left( \Omega_{\rm bg}^0 (1 + z_{\rm v})^3 + \Omega_\Lambda^0 \right)^{-\frac{1}{3}}. \tag{56}$$

Notice that the maximum virial radius allowed with non-vanishing $\Lambda$ is smaller than the maximum virial radius allowed without cosmological constant (57). The maximum virial radius reads

$$R_{\rm v}^{\rm max} \simeq 20 \left( \frac{M_{\rm v}}{10^{12} h^{-1} M_{\odot}} \right)^{\frac{1}{3}} \times \left( \Omega_{\rm bg}^{0} (1+z_{\rm v})^3 + \Omega_{\Lambda}^{0} \right)^{-\frac{1}{3}} h^{-1}{\rm Mpc}. \quad (57)$$

We can estimate the virial radius as $R_{\rm v} \simeq \mu R_{\rm v}^{\rm max}$ where the $\mu$ parameter is selected in such a way that the inequality (57) is satisfied; namely, we choose $\mu \simeq 10^{-2}$. Besides, we can verify that the second term in (55) is considerably small if $R_{\rm v} \simeq \mu R_{\rm v}^{\rm max}$. We know the observational value for the cosmological constant, $\Lambda = 1.105 \times 10^{-6}{\rm Mpc}^{-2}$, and we take as the virialized mass $M_{\rm v} = 10^{12} h^{-1} M_{\odot}$, so the ratio $\Lambda\mu^3 (R_{\rm v}^{\rm max})^3/3GM_{\rm v} \simeq 3.8 \times 10^{-3} \ll 1$ where $h = 0.7$ and $G$ being the Newton constant. The latter fact confirmed that idea that the cosmological constant introduces a small deviation in the dispersion velocity obtained from the virial theorem.

## 3 Summary

In this work, we analyzed the dynamics and the collapse of a collisionless self-gravitating system composed of dark matter and baryonic matter. This system is described by two Boltzmann equations, one for each component, gravitationally coupled through the Poisson equation. First, we derived the general solution for the total density perturbation and then we solved numerically the coupled master equation for both components in different cases, showing that for modes deep inside the Hubble horizon and under the condition $q \ll q_{J0}$ the density of matter tends to grow. In fact, we showed that a baryon perturbation catches up with the dark matter perturbation at late times, so dark matter is the driving force which provides the gravitational potential wells for the baryons to fall into, allowing them to create bound structures. Besides, we also examined in broad terms what happens with this toy model in the non-linear regime by working within the realm of Newtonian theory. We considered the formation of non-linear structures in the form of a spherical over-density of radius $R$ and mass $M$ embedded into the homogeneous spatially flat and matter-dominated universe. We explored the linear regime until the turn-around event followed by the collapse and the subsequently virialization process. The viral theorem along with the fact that the virialization criterion implies that galaxies were formed at low redshift, say less than ten. We also discussed the case with non-zero cosmological constant.

It is important to mention the main elements that we should introduce in the kinetic theory to study a composed system with three components (dark matter, baryons, and dark energy) within the Newtonian approach without going to the full relativistic formulation. We have basically three distribution functions with their corresponding Boltzmann equations along with one Poisson equation that couples gravitationally all the aforesaid components. We may consider the case of dark energy decoupled from the baryon plus dark matter system. The possibility of an extra-coupling between dark matter and baryons has been suggested by the Experiment to Detect the Global Epoch of Reionization Signature (EDGES) measuring the 21-cm absorption signal from primordial neutral hydrogen at redshift $z \simeq 17$ [22–24]. This signal is considerably stronger than what is expected from the vanilla cosmic model and contains potentially new information about the true nature of dark matter. If that is the case under study, the interaction between dark matter and baryons must be specified through the collision operator. Such an interaction will affect the structure formation due to the exchange of momentum, energy and flux of heat; however, the specific result will strongly depend on the collision operator and therefore the specific interactions considered (say decay, annihilation, etc.). In fact, an interaction could potentially suppress the growth of structure in the early universe. Of course, we could expect that the structure formation will be totally ineffective at late times due to the overall expansion of the universe. Besides, one possible extension of the current work is to consider also the kinetic theory with a collision operator that takes into account the interaction between dark matter and dark energy in order to explain the current excess of brightness in the 21 cm line as proposed by several authors in the literature [25–28]. We will address the latter possibility within our formalism in the near future.

**Acknowledgements** G.M.K. is supported by the Conselho Nacional de Desenvolvimento Científico e Tecnológico (CNPq)-Brazil. M.G.R. is supported by a FAPES/CAPES Grant under the PPGCosmo Fellowship Programme. E.M.S. is supported by the Coordenação de Aperfeiçoamento de Pessoal de Nível Superior (CAPES)-Brazil.

**Data Availability Statement** This manuscript has no associated data or the data will not be deposited. [Authors' comment: This is a theoretical research work so no experimental data have been considered.]

Funded by SCOAP$^3$.